# Applications of Dynamical Systems in Engineering

*Yousuf Ibrahim Khan*

*Department of Electrical and Electronic Engineering*

*American International University-Bangladesh*

**Abstract :** This paper presents the current possible applications of Dynamical Systems in Engineering. The applications of chaos, fractals have proven to be an exciting and fruitful endeavor. These applications are highly diverse ranging over such fields as Electrical, Electronics and Computer Engineering. Dynamical Systems theory describes general patterns found in the solution of systems of nonlinear equations. The theory focuses upon those equations representing the change of processes in time. This paper offers the issue of applying dynamical systems methods to a wider circle of Engineering problems. There are three components to our approach: ongoing and possible applications of Fractals, Chaos Theory and Dynamical Systems. Some basic and useful computer simulation of Dynamical System related problems have been shown also.

**Keywords:** Dynamical Systems, Chaos, Fractals, Control, Feedback.

**1. Introduction :** The general forms for dynamical systems are,

x′ = f(x) [continuous time]

x(k+1) = f(x(k)) [Discrete time]

$\frac{dx}{dt} = ax$ is the simplest differential equation. Here x = x(t) is an unknown real valued function of a real variable **t** and $\frac{dx}{dt}$ is it's derivative; we can also use x′ or x′(t) for the derivative. For each value of parameter **a** we have a different differential equation. So for every value of t,

x′(t) = ax(t); to solve it, since x(t) = $ke^{at}$ is a solution so, x′(t)= $ake^{at}$ =ax(t)

we call the collection of all solutions of a differential equation the general solution of the equation. From the initial value problem,

$$x'=ax, \quad x(0) = u_0 \qquad (1)$$

The equation x' = ax is stable in a certain sense if a ≠ 0. If **a** is replace by another constant **b** whose sign is the same as **a,** then the qualitative behavior of the solutions doesn't change. But if a=0, the slightest change in **a** leads to a radical change in the behavior of solutions. We say, bifurcation at a=0 in the one parameter family of equations x'=ax. A system of differential equations is a collection of **n** interrelated different equations of the form,

$$x'_1 = f_1(t, x_1, x_2, ..., x_n) \qquad (2)$$

$$x'_2 = f_2(t, x_1, x_2, ..., x_n) \qquad (3)$$

$$x'_3 = f_n(t, x_1, x_2, ..., x_n) \qquad (4)$$

This can be written, x' = F(t, x); a solution would be of the form $X(t)=(x_1(t),...,x_n(t))$ that satisfies the equation so that,

$$X'(t)=F(t, X(t)) \qquad (5)$$

The system of equations is called autonomous if none of the $f_j$ depends on t, so the system becomes x' = F(x). A vector $X_0$ for which $F(X_0)=0$ is called an equilibrium point for the system. An equilibrium point corresponds to a constant solution $X(t) \equiv X_0$ of the system. Many of the most important differential equations encountered in engineering are second order differential equation, like x" = f(t,x,x')

Any second order system may be written as a system of first order equations. Unlike linear (constant coefficient) systems, where we can always find the explicit solution of any initial value problem, this is rarely the case for nonlinear systems.

In general, a dynamical system is a mapping from the space of input signals to the space of output signals. By the term signal, we mean a real vector-valued function of a time variable. For an arbitrary dynamical system there are two major kinds of stability notions: internal stability and external stability. The internal stability considers trajectories of an autonomous system, i.e. system without any inputs and outputs; this way, it is a property of internal dynamics of the system. External stability concerns with how much the system amplifies signals. Everything in nature is continuously changing and evolving. Any system whose status changes with time is called a dynamical system. Dynamical Systems are described by differential equations-whose solution show how the variables of the system depend on the independent variable time.

Roughly speaking, a dynamical system describes the evolution of a state over time. To make this notion more precise we need to specify what we mean by the terms "evolution", "state" and "time". Certain dynamical systems can also be influenced by external inputs, which may represent either uncontrollable disturbances (e.g. wind affecting the motion of a wind turbine) or control signals (e.g. the commands of the programmer to the a robotic spacecraft for controlling engines). Some dynamical systems may also have outputs, which may represent either quantities that can be measured, or quantities that need to be regulated. Dynamical systems with inputs and outputs are sometimes referred to as control systems which is a very important topic in Engineering.

Based on the type of their state, dynamical systems can be classified into:

1. Continuous 2. Discrete 3. Hybrid

Based on the set of times over which the state evolves, dynamical systems can be classified into:

1. Continuous time 2. Discrete time 3. Hybrid time

Continuous state systems can be further classified according to the equations used to describe the evolution of their state:

1. Linear 2. Nonlinear

A dynamical system consists of two parts:

1. State 2. Dynamics

The state is defined as the information necessary at a given time instant to define further outputs of the system. The dynamics are a set of rules that define how the state evolves over time. A dynamical system evolves on a state space (also called phase space). The state is represented by a vector in the state space.

A mathematical description of a dynamical system is

1. System of difference equations for the discrete case

2. Systems of differential equations for the continuous case

The motion of the system is called a trajectory or orbit of the dynamical system.

The revolution in digital technology has fueled a need for design techniques that can guarantee safety and performance specifications of embedded systems, or systems that couple discrete logics with analog physical environment. Such systems can be modeled by hybrid systems, which are dynamical systems that combine continuous-time dynamics modeled by differential equations and discrete-event dynamics modeled by finite automata. Important applications of hybrid systems include CAD, real-time software, robotics and automation, mechatronics, aeronautics, air and ground transportation systems, process control etc. Recently, hybrid systems have been at the center of intense research activity in the control theory, computer-aided verification, and artificial intelligence communities, and methodologies have been developed to model hybrid systems, to analyze their behaviors, and to synthesize controllers that guarantee closed-loop safety and performance specifications.

Hybrid systems are dynamical systems that involve the interaction of different types of dynamics. Hybrid dynamics provide a convenient framework for modeling systems in a wide range of Engineering Applications :

a. In electrical circuits continuous phenomena such as the charging of capacitors are interrupted by discrete phenomena such as by switching opening and closing, or diodes going on or off.

b. In embedded computation systems a digital computer interacts with a mostly analog environment

Nonlinear, chaotic systems can produce very irregular data; similar but distinct from a stochastic system (i.e. system with probabilistic dynamics).

1. The rapid loss of predictability of chaotic systems is due to a phenomenon called sensitive dependence on initial conditions.

2. For a certain class of dynamical systems, once transients are over, the trajectory of the system approaches a subset of the state space called an attractor.

3. Attractors of dissipative chaotic systems generally are strange attractors-complicated geometrical objects that exhibit fractal structure.

4. Examples of deterministic chaotic systems: Logistic Map, Lorenz attractor.

In a linear system, there can be only one equilibrium point, and the structure of the vector field over the whole state space is same-determined by the Eigenvalues and Eigenvectors of the Matrix. So Linear Systems are simple to Analyze. But actually all systems are practically Nonlinear. In general, a linear set of equations is actually a local linear approximation of a nonlinear system in the neighborhood of an equilibrium point.

A dynamical system is a way of describing the passage in time of all points of a given space **S**. **S** could be thought of the space of states of some physical system. Mathematically **S** might be a Euclidean space or an open subset of Euclidean space or some other space such as a surface in $\mathbb{R}^n$. Given an initial position X $\in \mathbb{R}^n$, a dynamical system on $\mathbb{R}^n$ tells us where X is located 1 unit of time later, 2 units time later, and so on. At time zero, X is located at position X. If we measure the positions $X_t$ using only integer time values, we have a discrete dynamical system. If time is measured continuously with $t \in \mathbb{R}$, we have a continuous dynamical system. If the system depends on time in a continuously differential manner, we have a smooth dynamical system.

Chua's circuit is an RLC circuit for the study of chaos with four linear elements and a nonlinear diode, which can be modeled by a system of three differential equations. The equations for Chua's are,

$$X' = c_1(y - x - g(x)) \tag{6}$$

$$Y' = c_2(x - y + z) \tag{7}$$

$$Z' = -c_3 y \tag{8}$$

Where, $g(x) = m_1 x + \dfrac{m_0 - m_1}{2}(|x+1| - |x-1|)$

The function g(x) is the only nonlinearity in the circuit. From the simulation we can see attractors from this circuit. The term attractor is used for the forward time limit of an orbit that attracts a significant portion of initial conditions. A sink is an example since it attracts at least a small neighborhood of initial values. Chaotic orbits can be attracting, If the forward limit set of such a chaotic orbit contains the orbit itself (and therefore contains a dense orbit), then the attractor is a chaotic attractor.

In Fractals, If we consider a bounded set A in a Euclidean n-dimensional space, then the set A is said to be self-similar if A is the union of N distinct ( non overlapping) copies of itself, each of which has been scaled by a ration r < 1 in all coordinates. The fractal is described by the relationship, $Nr^D = 1$ or $D = -\dfrac{\ln N}{\ln r}$, where D is the fractal dimension. The ranges in the value of D characterize the type of fractal. Fractals are mathematical objects which can be in general be written as,

$$A = \lim_{n \to \infty} \tau^n (A_0) \tag{9}$$

Where $A_0$ denotes an initial object and $\tau^n = \tau * \tau * \tau \ldots \tau$ (n times)

Denotes n iteration of $\tau$. They are typically generated by computing and displaying a sequence of iterates, $A_0, A_1, A_2, \ldots$

Where, $A_n = \tau (A_{n-1})$. It is clear that fractals satisfy the invariance equation:

$$A = \tau (A) \tag{10}$$

Which confers to them a property which we generally refer to as self-transformability, and which leads to the well known properties of fractals to be rugged objects with an infinite amount of detail, objects which can be found again and again in magnified small pieces of themselves.

## 2. Fractals in Engineering :

**2.1. Fractal Image Compression :** Storing images in less memory leads to a direct reduction in storage cost and faster data transmissions which is important in both signal processing and communication engineering. These facts justify the efforts, of companies and universities, on new image compression algorithms. Images are stored on computers as collections of bits (a bit is a binary unit of information which can answer "yes" or "no" questions) representing pixels or points forming the picture elements. Since the human eye can process large amounts of information (some 8 million bits), many pixels are required to store moderate quality images. These bits provide the "yes" and "no" answers to the 8 million questions that determine the image. Most data contains some amount of redundancy, which can sometimes be removed for storage and replaced for recovery, but this redundancy does not lead to high compression ratios. An image can be changed in many ways that are either not detectable by the human eye or do not contribute to the degradation of the image [1]. The standard methods of image compression come in several varieties. The current most popular method relies on eliminating high frequency components of the signal by storing only the low frequency components (Discrete Cosine Transform Algorithm). This method is used on JPEG (still images), MPEG (motion video images), H.261 (Video Telephony on ISDN lines), and H.263 (Video Telephony on PSTN lines) compression algorithms.

Fractal Compression was first promoted by M.Barnsley, who founded a company based on fractal image compression technology but who has not released details of his scheme. The first public scheme was due to E.Jacobs and R.Boss of the Naval Ocean Systems Center in San Diego who used regular partitioning and classification of curve segments in order to compress random fractal curves (such as political boundaries) in two dimensions [2]. A doctoral student of Barnsley's, A. Jacquin, was the first to publish a similar fractal image compression scheme [3].

Imagine a special type of photocopying machine that reduces the image to be copied by half and reproduces it three times on the copy (see Figure 1). What happens when we feed the output of this machine back as input? Figure 2 shows several iterations of this process on several input images. We can observe that all the copies seem to converge to the same final image, the one in 2(c). Since the copying machine reduces the input image, any initial image placed on the copying machine will be reduced to a point as we repeatedly run the machine; in fact, it is only the position and the orientation of the copies that determines what the final image looks like.

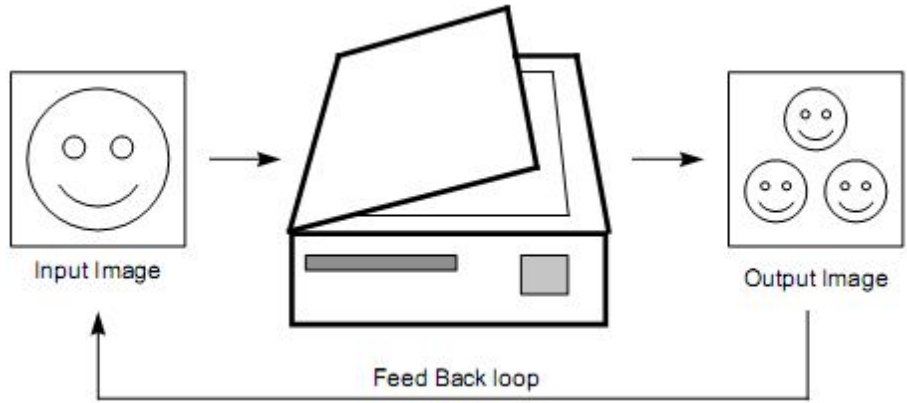

Fig 1 : A copy machine that makes three reduced copies of the input image

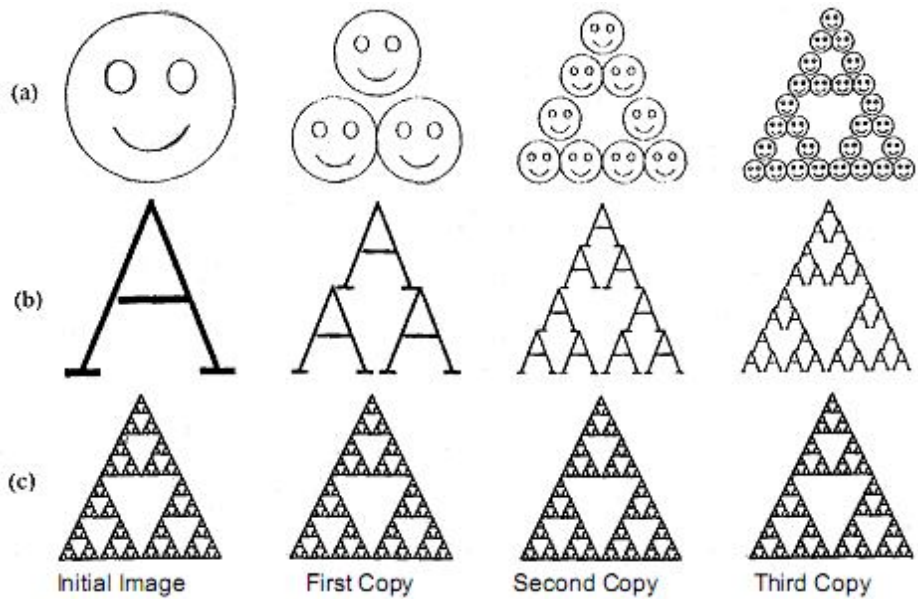

Fig 2 : The first three copies generated on the copying machine Figure 1.

The way the input image is transformed determines the final result when running the copy machine in a feedback loop. However we must constrain these transformations, with the limitation that the transformations must be contractive, that is, a given transformation applied to any two points in the input image must bring them closer in the copy. This technical condition is quite logical, since if points in the copy were spread out the final image would have to be of infinite size. Except for this condition the transformation can have any form.

A common feature of these transformations that run in a loop back mode is that for a given initial image each image is formed from a transformed (and reduced) copies of itself, and hence it must have detail at every scale. That is, the images are fractals. This method of generating fractals is due to John Hutchinson [5], and more information about the various ways of generating such fractals can be found in books by Barnsley [4] and Peitgen, Saupe, and Jurgens. Barnsley suggested that perhaps storing images as collections of transformations could lead to image compression.

The scheme will encode an image as a collection of transforms that are very similar to the copy machine metaphor. Just as the fern has detail at every scale, so does the image reconstructed from the transforms. The decoded image has no natural size, it can be decoded at any size. The extra detail needed for decoding at larger sizes is generated automatically by the encoding transforms. One may wonder if this detail is "real"; we could decode an image of a person increasing the size with each iteration, and eventually see skin cells or perhaps atoms. The answer is, of course, no. The detail is not at all related to the actual detail present when the image was digitized; it is just the product of the encoding transforms which originally only encoded the large-scale features. However, in some cases the detail is realistic at low magnifications, and this can be useful in Security and Biomedical Engineering Imaging applications.

A typical image of a face, does not contain the type of self-similarity like the fern. The image does contain other type of self-similarity. Like, if someone stands in front of the mirror outside and takes an image then the reflection portion on face from sunlight is same as the portion that we get from the mirror. These distinctions form the kind of self-similarity rather than having the image be formed by whole copies of the original (under appropriate affine transformations), here the image will be formed by copies of properly transformed parts of the original. These transformed parts do not fit together, in general, to form an exact copy of the original image, and so we must allow some error in our representation of an image as a set of transformations.

**2.2. Fractal Antenna :** Classical geometry limits the packing of large wave-lengths into small devices. By leveraging the space-filling properties of fractals, it is possible to extract the maximum performance from miniature antennas. The fractal repetition of patterns at reducing scales also enables researchers to design multiband antennas with the highest possible density of bands.

Fractal antenna theory is built, as is the case with conventional antenna theory, on classic electromagnetic theory.  Fractal antenna theory uses a modern (fractal) geometry that is a natural extension of Euclidian geometry. A fractal antenna is an antenna that uses a fractal, self-similar design to maximize the length, or increase the perimeter (on inside sections or the outer structure), of material that can receive or transmit electromagnetic radiation within a given total surface area or volume. Such fractal antennas are also referred to as multilevel and space filling curves, but the key aspect lies in their repetition of a motif over two or more scale sizes [6] or "iterations". For this reason, fractal antennas are very compact, multiband or wideband, and have useful applications in cellular telephone and microwave communications.

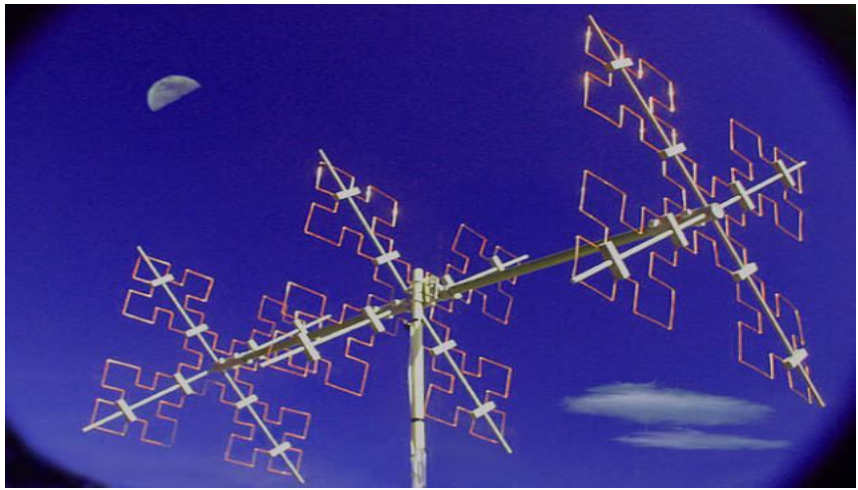

Fig 3: An example of Fractal Antenna

A good example of a fractal antenna as a space filling curve is in the form of a shrunken fractal helix. Here, each line of copper is just a small fraction of a wavelength. A fractal antenna's response differs markedly from traditional antenna designs, in that it is capable of operating with good-to-excellent performance at many different frequencies simultaneously. Normally standard antennas have to be "cut" for the frequency for which they are to be used—and thus the standard antennas only work well at that frequency. This makes the fractal antenna an excellent design for wideband and multiband applications.

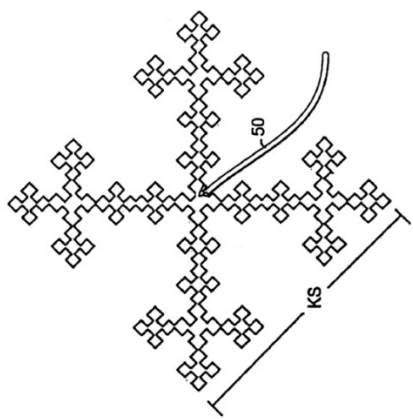

Fig 4 : An example of a fractal antenna: a space-filling curve called a Minkowski Island

Many fractal element antennas use the fractal structure as a virtual combination of capacitors and inductors. This makes the antenna so that it has many different resonances which can be chosen and adjusted by choosing the proper fractal design. Just like any other antenna, the physical size of the antenna is related to its potential bandwidth, and the resonant frequency changes depending on the amount of fractal reactive loading. In this manner, the fractal antenna, just like any other antenna with reactive loading (i.e. dielectric, ferrites, capacitors, inductors, etc.), can have its resonant frequency lower than that of the typical free-space half-wavelength fundamental resonant frequency predicted by setting the largest physical dimension of the antenna equal to half a wavelength. In general, although their effective electrical length is longer, the fractal element antennas are themselves physically smaller, again due to reactive loading.

Fractal element antennas are shrunken compared to conventional designs, and do not need additional components, assuming the structure happens to have the desired resonant input impedance. In general the fractal dimension of a fractal antenna is a poor predictor of its performance and application. Not all fractal antennas work well for a given application or set of applications. Computer search methods and antenna simulations are commonly used to identify which fractal antenna designs best meet the need of the application.

Although the first validation of the technology was published as early as 1995, recent independent studies show advantages of the fractal element technology in real-life applications, such as RFID [8] and cell phones [9].

Antenna theory considers three classes of radiators in terms of frequency coverage: (1) narrowband – small range of the order of a few percent around the designed operating frequency, (2) wideband or broadband – covers an octave or two, and (3) frequency independent – about a ten to one or greater range of frequencies. Any good antenna text talks about antenna scaling, that is the properties (impedance, efficiency, pattern, etc.) remain the same if all dimensions and the wavelength are scaled by the same factor.

Now, remembering that a fractal is a figure that "looks" the same independent of size scaling, we come upon the amazing realization that a fractal shaped metal element can be used as an antenna over a very large band of frequencies. A typical book would say something like, "A distinguishing feature of frequency independent antennas is their self-scaling behavior". But then go on to say, "Frequency independent antennas can be divided into two types: spiral antennas and log-periodic antennas". To remedy this situation a large volume of research has been published on various aspects of fractal antennas and fractal electromagnetics. It is interesting to note that, as fractal geometry is a superset of Euclidian geometry, so is fractal (geometry based) antenna theory a superset of classic (Euclidian geometry) antenna theory. It is somewhat poetic that because of this set to superset relationship, fractal antenna analysis picks up (where classic theory lets off) with the spiral and the log-periodic structures. We are seeing fractal antenna theory shedding new light on our understanding of classic wideband antennas [7].

A fractal antenna is created using fractal geometry, a self-similar pattern built from the repetition of a simple shape. The inherent qualities of fractals enable the production of high performance antennas that are typically 50 to 75 percent smaller than traditional antennas. Fractal antennas are also reliable and cost-effective. Antenna performance is attained through the geometry of the conductor, rather than with the accumulation of separate components or elements that increase complexity and potential failure points. Fractal antennas also allow for multiband capabilities, decreased size, and optimum smart antenna technology. Fractal antennas can be produced in all existing antenna types, including dipole, monopole, patch, conformal, bicone, discone, spiral, and helical. Many hybrid designs greatly extend frequency ranges. The key benefits of fractal antenna technology are:

a) Reduced antenna size b) Multi-band functionality c) Improved antenna performance

The geometry of the fractal antenna encourages its study both as a multiband solution and also as a small (physical size) antenna. First, because one should expect a self-similar antenna (which contains many copies of itself at several scales) to operate in a similar way at several wavelengths. That is, the antenna should keep similar radiation parameters through several bands. Second, because the space-filling properties of some fractal shapes (the fractal dimension) might allow fractal shaped small antennas to better take advantage of the small surrounding space.

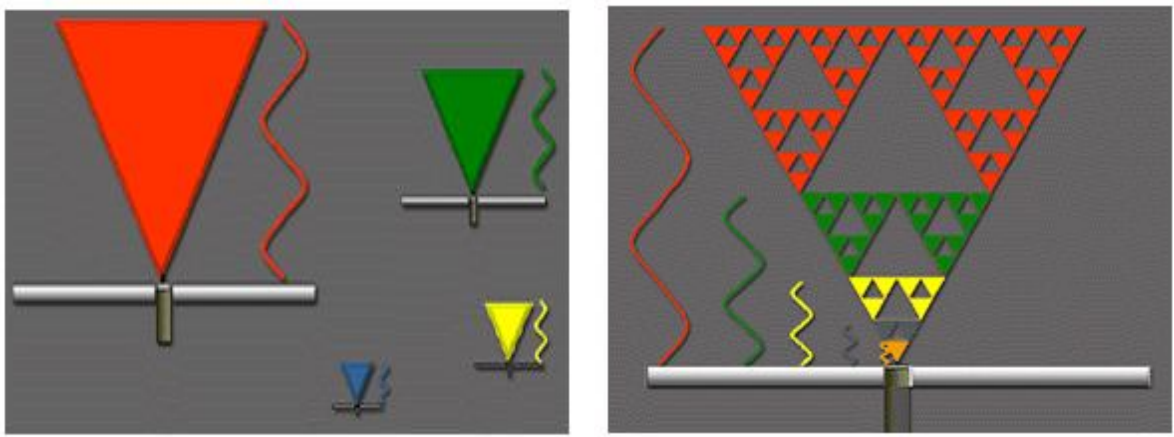

Fig 5 : Four antennas (with a wave cartoon) intended to be used on four discrete frequency bands (Left); One antenna (Sierpinski's Fractal) intended to be use for four discrete frequency bands (Right).

Also the study of the fractal random array remains one of the most interesting and fruitful areas in fractal antennas. The fractal random array utilizes a balance of long-range order (typical of fractals) and short-range disorder (typical of random arrays). The randomness provides robustness to element failure while the fractal or periodic structure provides the needed multiband or wideband performance.

Researchers now also have developed a new metamaterial technology that uses fractals to make layered, partless antennas and related electronics. Metamaterials are composites with unusual properties not found in nature. These new antennas, called metacloak antennas, have unique performance abilities in bandwidth, gain, directivity and versatility of form factor. Previously antennas had attributes in many form factor and performance regimes, but these new antennas are unprecedented. The new antennas have layers that are partless, with no electrical connections, are lightweight, have no ferrite or exotic materials, and are easy to make and implement. They can be far smaller and far thinner. Built up as layers of separated printed circuits to form a covering or 'cloak', the technology also applies to other EM spectral regimes, in addition to RF. The cloaking layers use close-spaced fractal resonators, tiny part less circuits, to accomplish the effect, and the scientific community has commonly called close-spaced resonators metamaterials. As the smaller size and pleasing bandwidth attributes of fractal resonators enable the new metamaterial advantages. The technology has also led to fractal-based arrays that are far smaller than expected from their gain, while maintaining very broadband ability. The new antenna technology is expected to allow antennas to go in many places they haven't been used before, especially in conformal or hidden platforms. It can be used to integrate antenna onto surfaces used for other things, thereby making the separate antenna platform notion a relic in many cases. Fractal sees the new technology as a logical extension of its core technologies and a new and versatile addition in solving challenging problems in electromagnetics.

**2.3. Fractal Capacitor :** Capacitors are one of the crucial elements in integrated circuits and are used extensively in many IC applications such as data converters, sample and holds, switched capacitor circuits, RF oscillators, and signal mixers. A high-density capacitor structure using fractal geometries that can be built in standard digital processes. The linearity of this structure is similar to the conventional parallel-plate metal to metal capacitor. The bottom plate parasitic capacitance of this structure is small, which makes it appealing for many circuit applications such as switched capacitor systems. Unlike conventional metal-to-metal capacitors, the density of a fractal capacitor increases with scaling. In this structure a lateral flux capacitor is used where two terminals of the device are built using a layer of metal, unlike a vertical flux capacitor, where two different metal layers must be used.

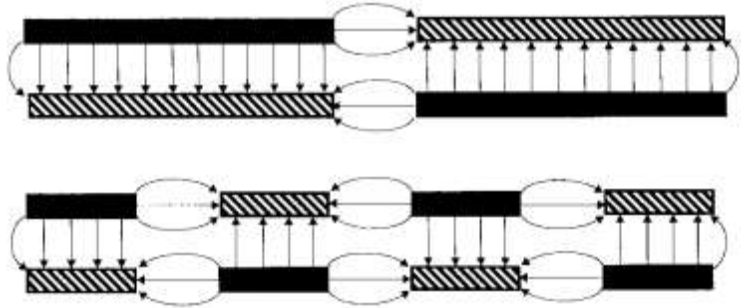

Fig : 6 Cross connected metal layers in Lateral flux capacitor (Capacitance is increased)

The increase in capacitance due to fringing is proportional to the periphery of the structure. Therefore, structures with large periphery per unit area like Fractals are desirable. So It is possible to increase the capacitance density of parallel-plate capacitors by exploiting lateral fringing fields in cross connected metal layers [10]. Fractals are geometric patterns that are repeated at ever smaller scales to produce irregular shapes and surfaces. They are geometrical structures with large perimeters. These suggest that fractals are good candidates for use in lateral flux capacitors. Although lithography limitations prevent fabrication of a real fractal, one can use quasi fractal geometries with feature sizes limited by lithography.

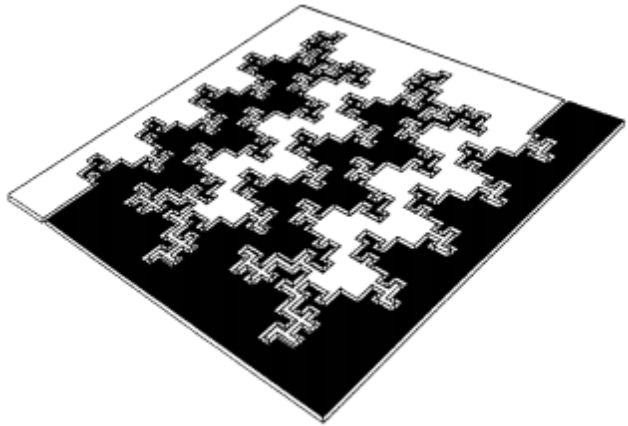

Fig: 7 Three Dimensional representation of a fractal capacitor using a single metal layer
(Fractals with large dimensions lead to greater capacitance)

This capacitor uses only one metal layer with a fractal border. The terminals of this square shaped capacitor have been identified using two different colors. One advantage of using lateral flux capacitors in general,

And fractal capacitors in particular, is the reduction of the bottom-plate capacitance. This reduction is due to two reasons. First, the higher density of the fractal capacitor (compared to a standard parallel-plate structure) results in a smaller area. Second, some of the field lines originating from one of the bottom plates terminate on the adjacent plate instead of the substrate, which further reduces the bottom-plate capacitance, as shown in fig.8. Because of this property, some portion of the parasitic bottom-plate capacitor is converted into the more useful plate-to-plate capacitance.

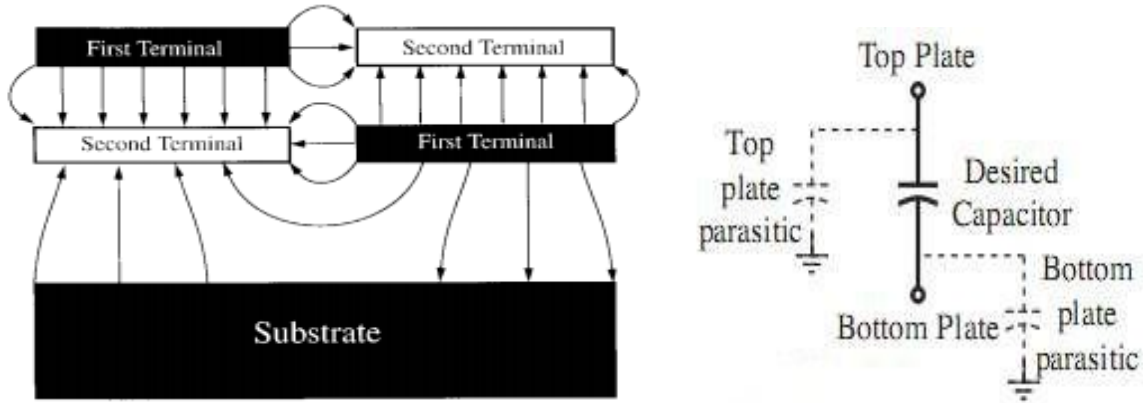

Fig: 8 Reduction of the bottom plate capacitance (Left); Parasitics are normally from the top and bottom plate to ac ground which is typically the substrate (Right)

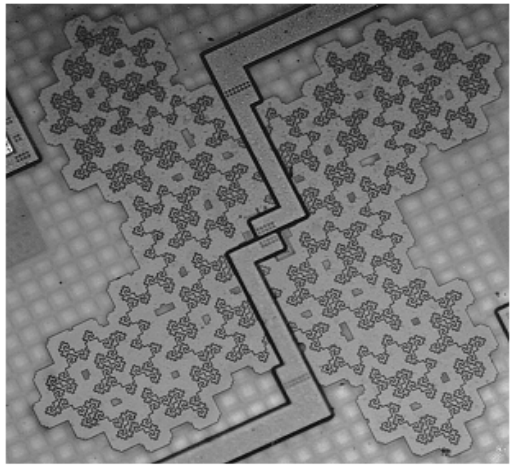

Fig 9: Die micrograph of a prototype fractal capacitor

This structure (fig. 9), exploits cross connected metal layers with a fractal border. Aside from the higher density, one advantage of the this structure is the reduction of bottom-plate capacitance due to the smaller area. In addition, since some of the field lines terminate on the adjacent plate instead of the substrate, the bottom plate capacitance is further reduced. The capacitance density of fractal structures increases with scaling of the process technologies, which makes them more attractive compared to standard parallel-plate capacitors.

**2.4. Fractal Filters, Resonators and Resistors:** One of the critical issues in the performance of a reactive component is the quality factor Q. The highest Q, the better the component. One of the critical issues in the performance of a reactive component is the quality factor Q. The highest the Q, the better the component. The Q is degraded by ohmic losses, but enhances with storage of reactive energy. In general terms, the performance of a small filter is related to the ability of the resonating structures of the filter in storing as much reactive energy as possible in the available volume. A good example of the energy storing capabilities of fractals is the Hilbert Resonator. Due to the unique space filling properties of fractals, a very long but small resonator can be efficiently packed into the same space as conventional half-wavelength resonator yet featuring a Q factor which is about 10 times larger.

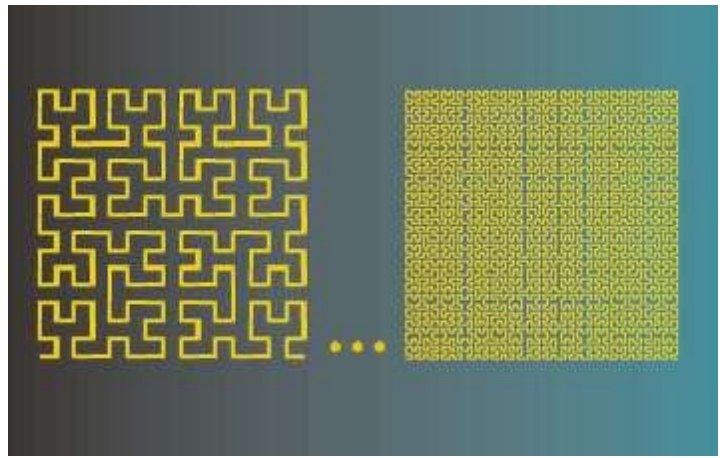

Fig 10 : The Hilbert Curve – Two different iterations. The curve grows exponentially at each iteration and completely fills up a square surface.

Due to the very long length of the strip, electromagnetic waves are bouncing back and forth and traveling for a longer time inside the resonator which results in a much higher stored energy. In general, in any application where the wavelength is large and a high-density package integration is convenient, fractals provide the optimum packaging technology. Low Q, high value fractal resistors is another area of fractal applications (Fig 12). Since an arbitrary large length of resistive material can be packed in an arbitrarily small area (with the only limitation of the manufacturing resolution), resistance per unit surface is maximized. At the same time, the parasitic serial inductance is minimized due to the ability of the fractal shape of filling space while maximizing the distance with itself due to its avoiding geometrical property [11].

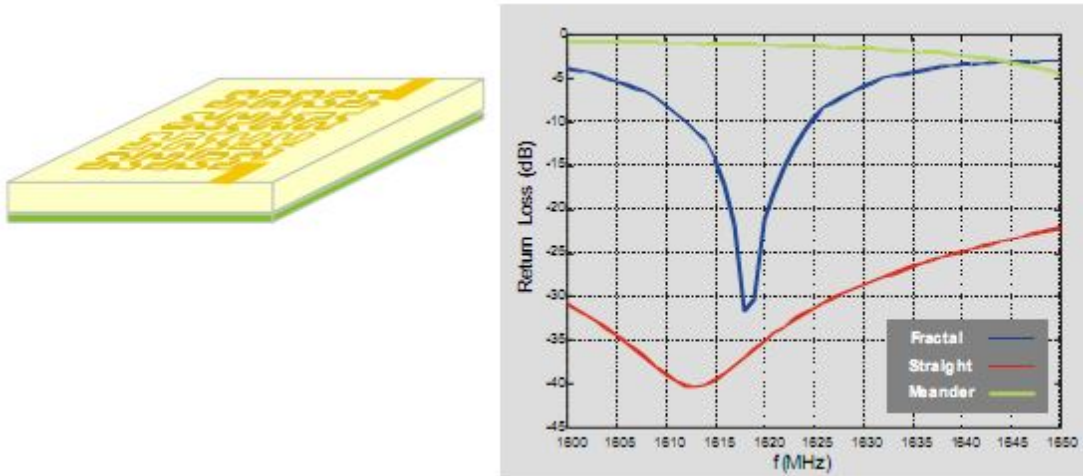

Fig 11 : The planar Hilbert Resonator is a high Q microstrip filter (Left); Comparison between a conventional half wavelength resonator, a meander line resonator and a Hilbert fractal resonator (Right)

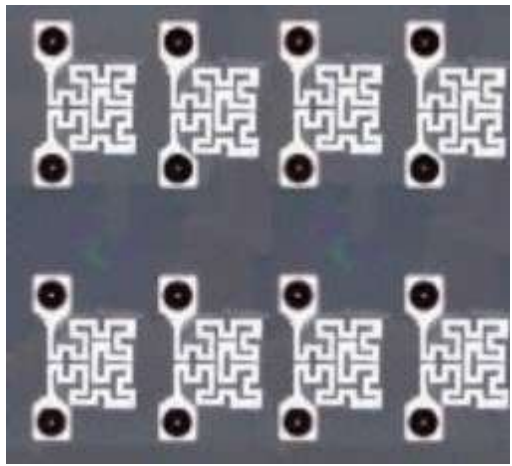

Fig 12 : Fractal Resistors

**2.5. Fractals in Nanotechnology :** 'Self-assembly' holds great promise as a technique for building commercial nano-circuits. Adopting this approach, the nano-engineer allows the circuit to build itself by exploiting natural growth processes. Self-assembly offers two striking advantages. Not only is it more efficient at assembling vast numbers of components compared to traditional fabrication techniques, this fundamentally 'green' technique constructs circuits by the addition of material rather than the wasteful removal of material that lies at the heart of previous 'top-down' fabrication techniques. One of the remarkable consequences of harnessing natural growth processes is that the resulting circuits exhibit natural patterns rather than the smooth, straight lines that form the framework of today's commercial circuit designs. In particular, many self-assembly processes generate fractal patterns. Fractals are shapes that repeat at many magnifications and are prevalent throughout nature, appearing in natural environments [12], biological systems and human physiology [13]. Nature uses fractals frequently because they possess a number of highly desirable properties. Topping this list is the fact that the repeating shapes build objects with huge surface areas. Nature exploits this property for example in trees, where the large surface area of the tree canopy ensures an unprecedented ability to absorb sunlight. The same approach could equally be employed to great effect by designing novel solar cell structures based on fractal shapes. Another consequence of large surface areas is that two merging patterns connect together very efficiently. For example, the dendritic structure of the neurons in the human brain exploits this fractal connectivity to produce enhanced information processing. The same connectivity could equally be exploited for future commercial computers by using artificial fractal electrical circuits. This philosophy of learning from nature's successes may well revolutionize many fields within nanotechnology. Although some electronics applications already exploit fractal geometry (cell phone antennae being a famous example), many fields lie at the start of this exciting journey, with many discoveries and challenges lying ahead. Current investigations focuses on two families of electronic device in which millions of metallic nano-particles (each approximately 50 nanometers across) are self-assembled into fractal circuits [16]. In the first family, the particles merge together to form 'nanoflowers'[14]using a growth process called diffusion-limited aggregation. In the second family, the nano-particles are attached to DNA strands [15] which assemble to form a fractal circuit. In both cases, the self-assembly process generates a tree-like pattern.

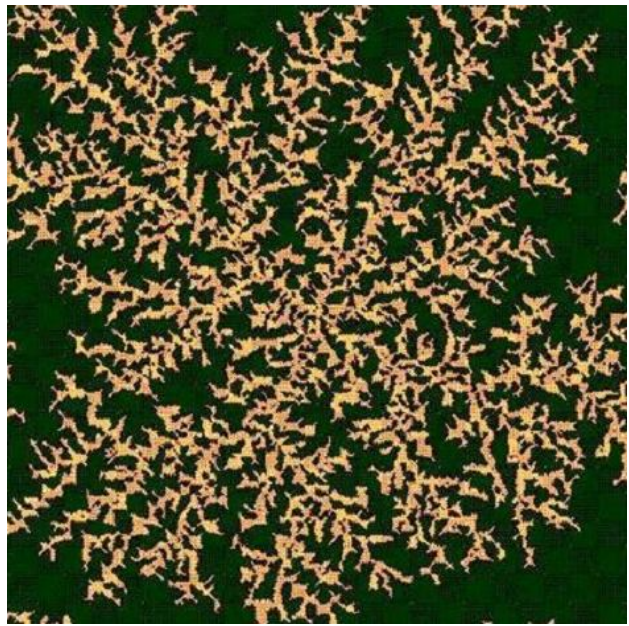

Fig 13 : Simulation of the self-assembled fractal electronic circuits

These projects are driven by the potential to tune the growth conditions so that the fractal characteristics of the circuits match those found for example in the neural structure of the human brain. Imagine a future where computers operate like our own minds and, ultimately, where fractal circuits may act as implants to be inserted into specific regions of the brain, restoring or enhancing a patient's mental functionality. Such goals represent the exceptional promise of nanotechnology - where researchers from a diverse range of disciplines work together to improve the basic quality of human life.

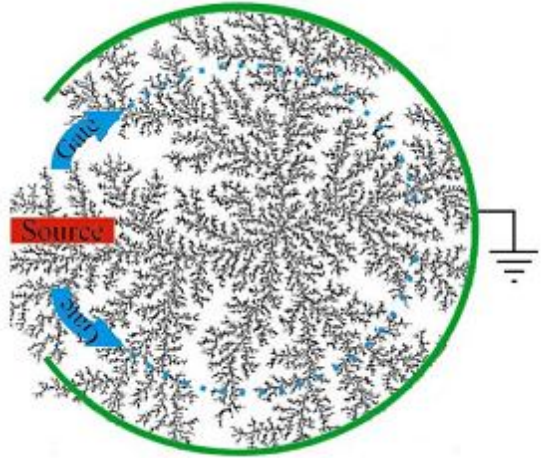

Fig 14: Fractal Field Effect Transistor (FET); schematic representation of a gating scheme for a DLA pattern

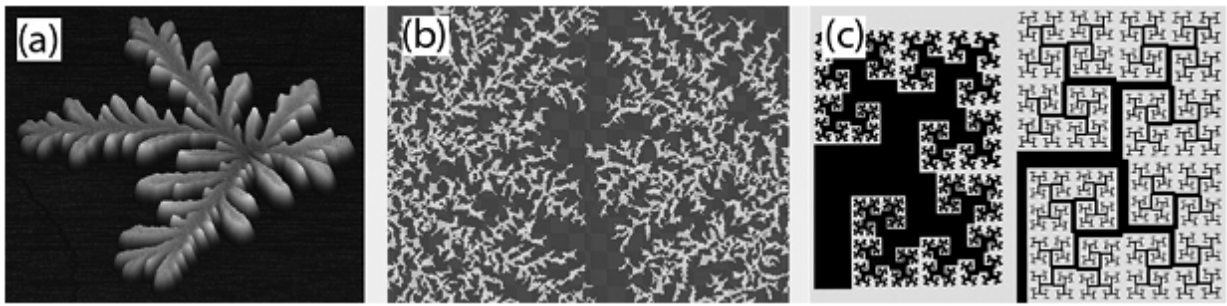

Fig15: (a) Fractal circuit formed using the diffusion-limited aggregation growth process
(b) Simulation of metallic fractal electrodes (light gray), photoactive material (dark gray)
(c) Examples of fractal branches with different scaling properties, course to fine ratios

**2.6 Other applications of Fractals :** The application of fractal geometry to speech signals and speech recognition systems is now receiving serious attention. A very important characteristic of fractals, useful for their description and classification, is their fractal dimension D. The fractal dimension provides an objective means of quantifying the fractal property of an object and comparing objects observed in the natural world. Fractals thus provide a simple description of many natural form. Intuitively D measures the degree of irregularity over multiple scales. Fractal speech recognition can be generally defined as the process of transforming continuous acoustic speech signals into a discrete representation. It involves the identification of specific words, by comparison with stored templates. The application of fractal geometry for modeling naturally occurring signals and images is well known. This is due to the fact that the statistics and spectral characteristics of random scaling fractals are consistent with many objects found in nature, a characteristic that is expressed in the term 'statistical self-affinity'.

## 3. Chaos in Engineering :

One interesting possibility opens up in systems of order three or greater : to get waveforms that do not have any periodicity. In such a case, the system state remains bounded-within a definite volume in the state space, but the same state never repeats. In every loop through the state space the state traverses a new path. This situation is called chaos and the resulting attractor is called a strange attractor. The system undergoes apparently random oscillations.

Even though long-term prediction may fail if a system is chaotic, an engineer need not be over-concerned about this failure. Rarely does an engineer need to predict the future state of a system so accurately. An engineer is more concerned with the overall properties of the orbit of a system. Even if one doesn't know the future state of the system, from the numerical solution of the concerned differential equations one can say with great confidence that the state will not run to infinity, will not collapse, and the state will be "somewhere" within a definite volume of the state space.

One of the utilities of chaos is that it can provide a framework for analyzing where on the spectrum between pure signal and pure noise, a data set might fall. Chaos is a type of signal, but can appear to be noise if not analyzed properly. Chaotic signals are irregular in time, but highly structured in phase space. Phase space embedding therefore provides a tool for visualizing the structure of chaotic signals, and for distinguishing chaos from noise. Furthermore, noise, by definition, is infinitely dimensional, whereas chaos is (relatively small) finite dimensional. Time series data can therefore be "unfolded" into higher dimensional space by sampling data points at fixed distances. A new data point will be created from a single time point and some integer number of steps ahead of that time point.

**3.1. Chaos based Cryptography :** Over the past decade, there has been tremendous interest in studying the behavior of chaotic systems. They are characterized by sensitive dependence on initial conditions, similarity to random behavior, and continuous broad-band power spectrum. Chaos has potential applications in several functional blocks of a digital communication system: compression, encryption and modulation. The possibility for self-synchronization of chaotic oscillations [17] has sparked an avalanche of works on application of chaos in cryptography. An attempt only to mention all related papers on chaos and cryptography in this short presentation will result in a prohibitively long list; and, therefore, we refer the reader to some recent work [18]. Despite a huge number of papers published in the field of chaos-based cryptography, the impact that this research has made on conventional cryptography is rather marginal. This is due to two reasons:

a) First, almost all chaos-based cryptographic algorithms use dynamical systems defined on the set of real numbers, and therefore are difficult for practical realization and circuit implementation.

b) Second, security and performance of almost all proposed chaos-based methods are not analyzed in terms of the techniques developed in cryptography. Moreover, most of the proposed methods generate cryptographically weak and slow algorithms.

Cryptography is generally acknowledged as the best method of data protection against passive and active fraud [19]. An overview of recent developments in the design of conventional cryptographic algorithms is given in [20].

Sensitive dependence on initial conditions is a very valuable property for cryptographic algorithms because one of the desired features of a cryptographic algorithm is that if the initial conditions used to encrypt data are changed by a small amount, one bit for instance, the encrypted text should be wildly different. For example a chaotic map like logistic map could be used because when r = 3.9, the logistic map exhibits chaotic behavior. A cipher is another name for a cryptographic algorithm [21]. The purpose of a cipher is to take unencrypted data, called plaintext, and produce an encrypted version of it, called the ciphertext. There are two types of ciphers : block ciphers and stream ciphers.

Keys increase the degree of security because well known, off the shelf, time tested algorithms can be used. An encryption pair is a key and the encryption system that the key is used with. Thus in an encryption pair, only the key has to be secret. Every encryption pair can be thought of as different key-less encryption algorithm. Chaotic maps and cryptographic algorithms (or more generally maps defined on finite sets) have some similar properties: sensitivity to a change in initial conditions and parameters, random-like behavior and unstable periodic orbits with long periods. Encryption rounds of a cryptographic algorithm lead to the desired diffusion and confusion properties of the algorithm. Iterations of a chaotic map spread the initial region over the entire phase space. The parameters of the chaotic map may represent the key of the encryption algorithm. An important difference between chaos and cryptography is that encryption transformations are defined on finite sets, while chaos has meaning only on real numbers. Moreover, for the time being, the notions of cryptographic security and performance of cryptographic algorithms have no counterpart in chaos theory (Fig 16).

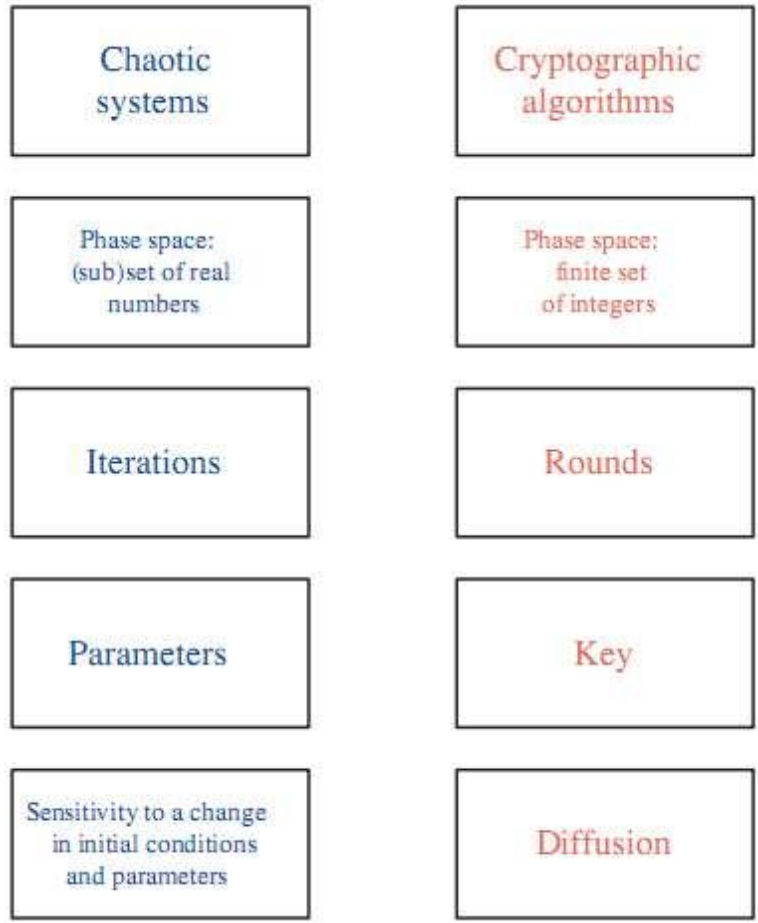

Fig 16: Similarities and differences between chaotic systems and cryptographic algorithms

Two general principles which guide the design of practical algorithms are diffusion and confusion [22]. Diffusion means spreading out of the influence of a single plaintext digit over many ciphertext digits so as to hide the statistical structure of the plaintext. An extension of this idea is to spread the influence of a single key digit over many digits of ciphertext. Confusion means use of transformations which complicate dependence of the statistics of ciphertext on the statistics of plaintext. The mixing property of chaotic maps is closely related to the property of diffusion in encryption transformations (algorithms). The keys of an encryption algorithm represent its parameters. Therefore, we should consider only such transformations in which both parameters and variables are involved in a sensitive way, that is "a small variation of any one" (variable, parameter) "changes the outputs considerably". In other words, a kind of "mixing property" should hold also in the parameter space of the map, if we would like to use chaotic maps as encryption algorithm. This implies that we consider only the maps for which chaos is persistent under small perturbations of parameters (keys).

**3.2. Chaotic Circuits :** The Chua circuit is among the simplest non-linear circuits that show most complex dynamical behavior, including chaos which exhibits a variety of bifurcation phenomena and attractors. In recent years chaos theory has attracted much interest in both the academic area and engineering study. One of the great achievements of the chaos theory is the application in secure communications. Chaotic signals depend very sensitively on initial conditions, have unpredictable features and noise like wideband spread spectrum. So, it can be used in various communication applications because of their features of masking and immunizing information against noise. The chaos communication fundament is the synchronization of two chaotic systems under suitable conditions if one of the systems is driven by the other. Chua's circuit is a simple oscillator circuit which exhibits a variety of bifurcations and chaos. The circuit contains three linear energy storage elements (an inductor and two capacitors), a linear resistor, and a single nonlinear resistor NR (Fig 17).

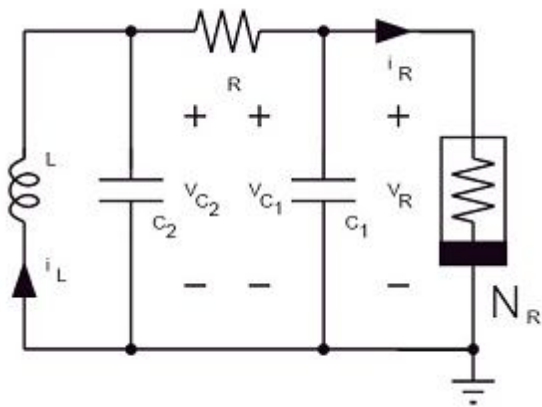

Fig 17: Schematic of Chua's Circuit. $N_R$ is the active nonlinear resistor, Called as the Chua diode.

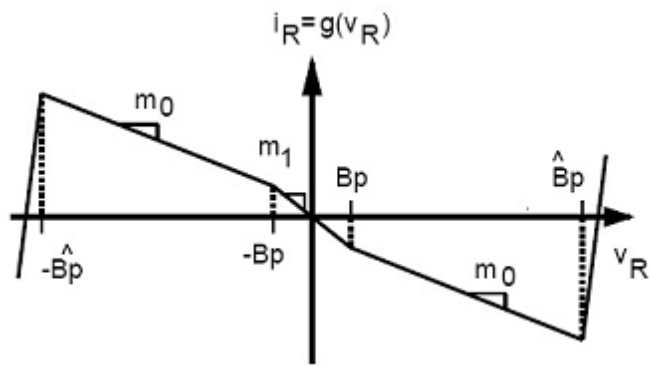

Fig 18 : Piece wise linear characteristic of the nonlinear resistor $N_R$ in Chua's circuit

Chua's circuit is the simplest autonomous (i.e no input signals) electronic circuit to generate chaotic signals. In fig.18 shows the typical piece wise linear characteristic of the non linear resistor $N_R$ (Chua Diode). Chua's circuit exhibits almost all known phenomenon of chaos. It has an easily implemented design and is a widely studied chaotic circuit.

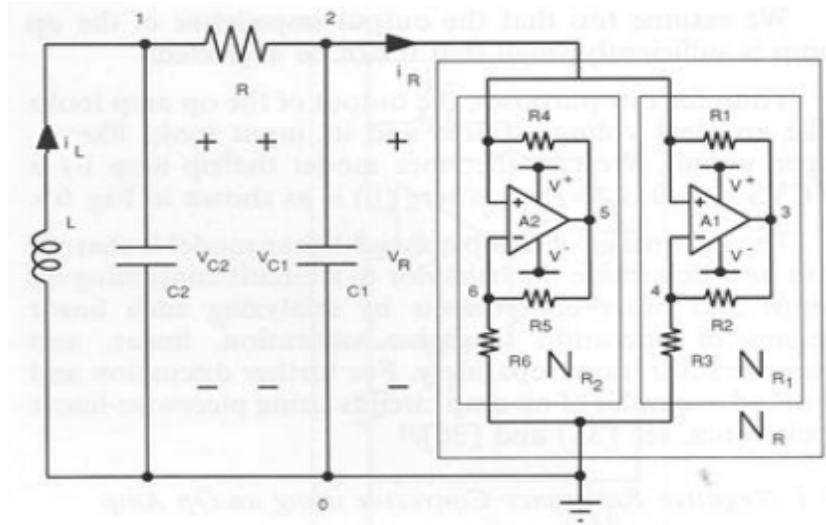

Fig : 19 Chua's circuit, the nonlinear resistor is realized using two operational amplifiers

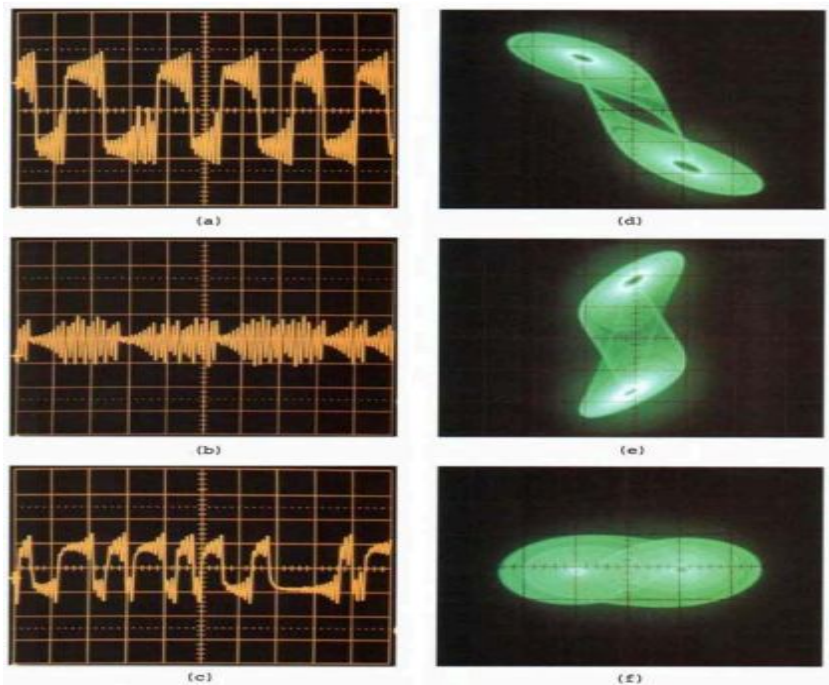

Fig 20: waveforms and phase plots recorded on an oscilloscope from experimental measures of Fig 3. The three waveforms in (a), (b) and (c) (left column) corresponds to $VC_1(t)$, $VC_2(t)$ and $i_L(t)$ from fig 3. The three phase plots displayed in (d), (e) and (f) correspond to the pair of variables ($VC_1$, $i_L$), ($VC_1$, $VC_2$) and ($VC_2$, $i_L$) respectively.

## 4. Dynamical Systems in Engineering :

**Feedback and Control Systems:** A dynamical system is a system whose behavior changes over time, often in response to external stimulation or forcing. The term feedback refers to a situation in which two (or more) dynamical systems are connected together such that each system influences the other and their dynamics are thus strongly coupled. Fig. 21 shows two feedback systems, one is open and another is closed. Feedback has many interesting properties that can be exploited in designing systems. It can also be used to create linear behavior out of nonlinear components, and a common approach in electronics. Feedback has many interesting and useful properties [23].

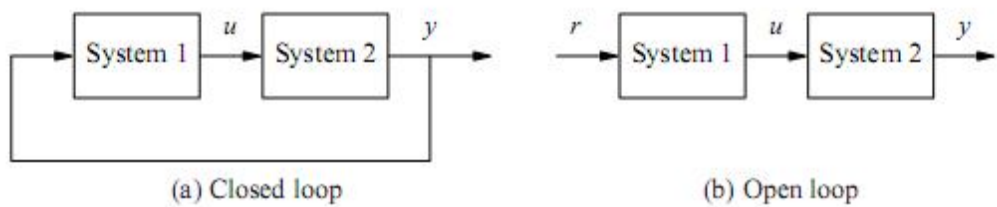

Fig : 21 Open and closed loop systems. (a) The output of system 1 is used as the input of systems 2, and the output of system 2 becomes the input of system 1, creating a closed loop system. (b) The interconnection between system 2 and system 1 is removed, and the system is said to be open loop.

It makes it possible to design precise systems from imprecise components and to make relevant quantities in a system change in a prescribed fashion. An unstable system can be stabilized using feedback, and the effects of external disturbances can be reduced. Feedback also offers new degrees of freedom to a designer by exploiting sensing, actuation and computation. The principle of feedback is simple: base correcting actions on the difference between desired and actual performance. Feedback can change the dynamics of a system. Through feed-back, we can alter the behavior of a system to meet the needs of an application, systems that are unstable can be stabilized, systems that are sluggish can be made responsive and systems that have drifting operating points can be held constants. Control theory provides a rich collection of techniques to analyze the stability and dynamic response of complex systems and to place bounds on the behavior of such systems by analyzing the gains of linear and non linear operators that describe their components. The term control has many meanings and often varies between communities. It means the use of algorithms and feedback in engineered dynamical systems. A modern controller senses the operation of a system, compares it against the desired behavior, computes corrective actions based on a model of the system response to external inputs and actuates the system to effect the desired change. This basic feedback loop of sensing, computation and actuation is the central concept in control. Control has had a major impact on electronics. The first application of feedback in electronics was patent on Vacuum tube amplifiers by the Robert Goddard in 1912, but the most influential development is undoubtedly the negative feedback amplifier (fig. 22). Negative feedback reduces the gain but makes the amplifier very insensitive to variations.

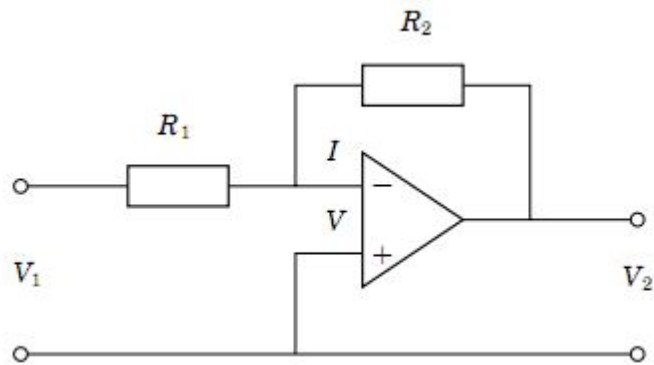

Fig: 22 An Amplifier with negative feedback

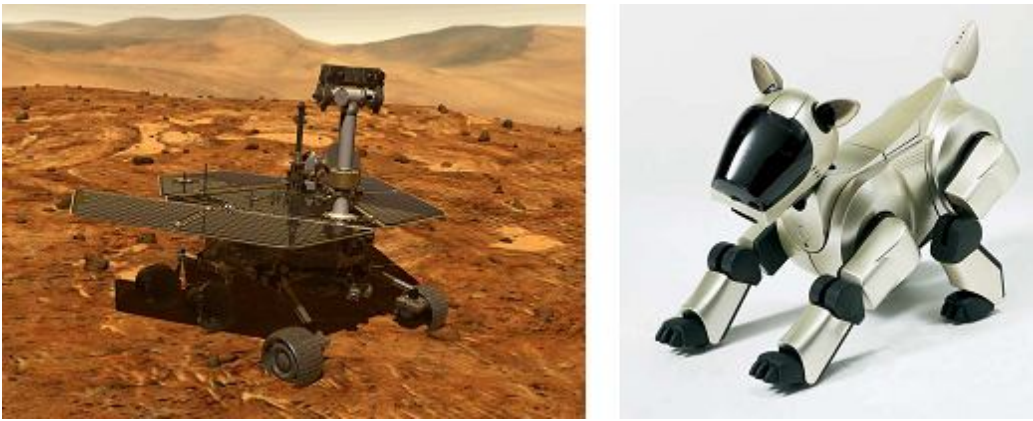

Fig : 23 Robotic systems; NASA (National Aeronautics and Space Administration) Mars Rover Spirit (Left) and Sony AIBO Entertainment Robot (Right). Both robots make use of feedback between sensors, actuators, and computation to function in unknown environments. These are also examples of Dynamical Systems.

The goal of cybernetic engineering has been to implement systems capable of exhibiting highly flexible or 'intelligent' responses to changing circumstances (Fig 23 : shows two robots). The higher level of feedback is a key element in robotics, where issues such as obstacle avoidance, goal seeking, learning and autonomy are prevalent.

# 5. Basic Simulations in Dynamical Systems

## (A) Graphical Iteration of the Logistic Map:

MATLAB Source Code :

```
fsize=15;
nmax=100;halfm=nmax/2;
t=zeros(1,nmax);t1=zeros(1,nmax);t2=zeros(1,nmax);
t(1)=0.2;
mu=3.8282;
axis([0 1 0 1]);
for n=1:nmax
    t(n+1)=mu*t(n)*(1-t(n));
end

for n=1:halfm
    t1(2*n-1)=t(n);
    t1(2*n)=t(n);
end

t2(1)=0;t2(2)=t(2);
for n=2:halfm
    t2(2*n-1)=t(n);
    t2(2*n)=t(n+1);
end
hold on
plot(t1,t2,'r');
fplot('3.8282*x*(1-x)',[0 1]);
x=[0 1];y=[0 1];
plot(x,y,'g');
hold off
%title('Graphical iteration for the tent map')
set(gca,'xtick',[0 1],'Fontsize',fsize)
set(gca,'ytick',[0 1],'Fontsize',fsize)
xlabel('x','Fontsize',fsize)
ylabel('f_{\mu}','Fontsize',fsize)
```

Simulation Result :

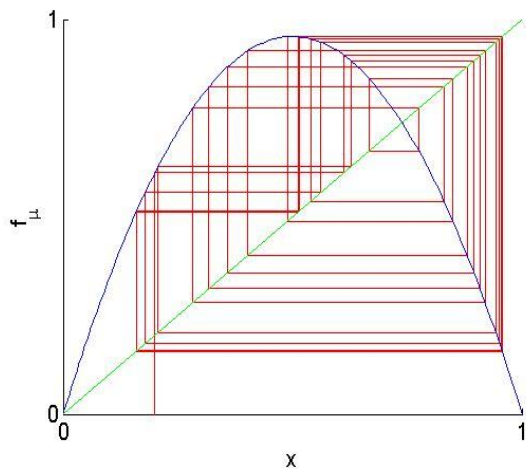

**(B) Chaotic Attractor of the Hénon Map :** The Hénon map is a discrete-time dynamical system. It is one of the most studied examples of dynamical systems that exhibit chaotic behavior.

MATLAB Source Code :

```
a=1.2;
b=0.4;
N=6000;
x=zeros(1,N);
y=zeros(1,N);
x(1)=0.1;
y(1)=0;
for n=1:N
    x(n+1)=1+y(n)-a*(x(n))^2;
    y(n+1)=b*x(n);
end
axis([-1 2 -1 1])
plot(x(50:N),y(50:N),'.','MarkerSize',1);
fsize=15;
set(gca,'XTick',-1:0.5:1,'FontSize',fsize)
set(gca,'YTick',-1:0.5:2,'FontSize',fsize)
xlabel('\itx','FontSize',fsize)
ylabel('\ity','FontSize',fsize)
```

Simulation Result:

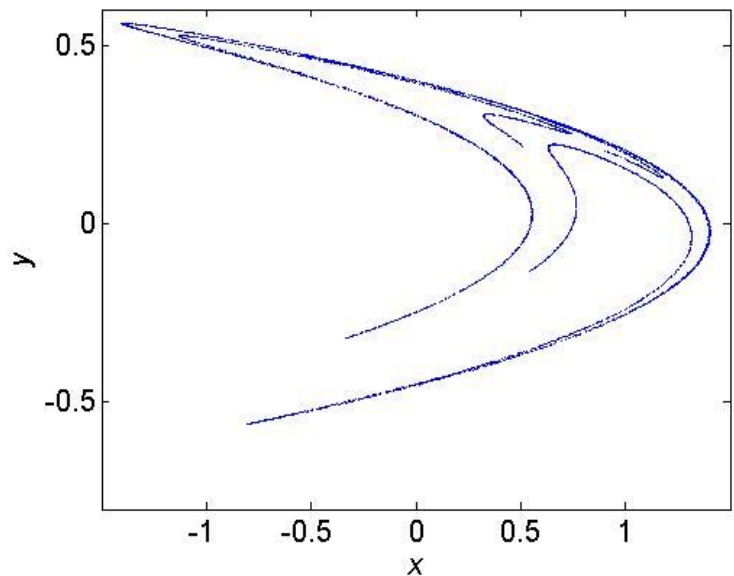

**(C) The Lorenz Attractor :**

MATLAB source code :

```
% Lorenz=inline('[10*(x(2)-x(1));28*x(1)-x(2)-x(1)*x(3);x(1)*x(2)-
(8/3)*x(3)]','t','x');
```

```
sigma=10;r=28;b=8/3;
Lorenz=@(t,x) [sigma*(x(2)-x(1));r*x(1)-x(2)-x(1)*x(3);x(1)*x(2)-b*x(3)];
options = odeset('RelTol',1e-4,'AbsTol',1e-4);
[t,xa]=ode45(Lorenz,[0 100],[15,20,30],options);
plot3(xa(:,1),xa(:,2),xa(:,3))
title('The Lorenz Attractor')

fsize=15;
xlabel('x(t)','Fontsize',fsize);
ylabel('y(t)','Fontsize',fsize);
zlabel('z(t)','FontSize',fsize);
```

Simulation Result :

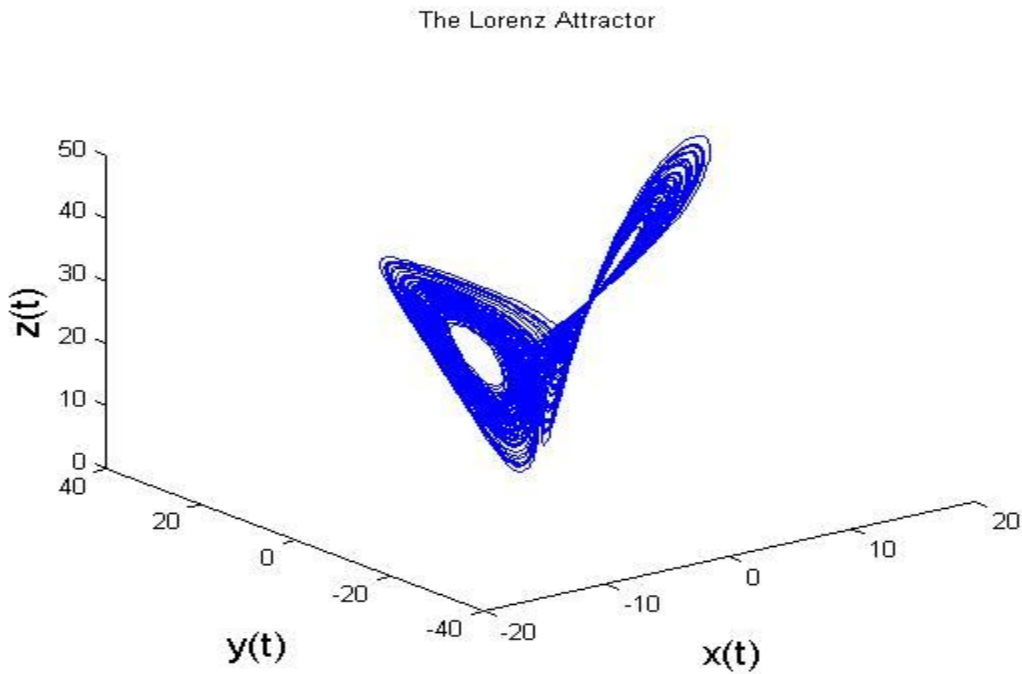

### (D) Chua's Attractor :

MATLAB source code :

```
Chua=@(t,x) [15*(x(2)-x(1)-(-(5/7)*x(1)+(1/2)*(-(8/7)-(-5/7))*(abs(x(1)+1)-
abs(x(1)-1))));x(1)-x(2)+x(3);-25.58*x(2)];
options = odeset('RelTol',1e-4,'AbsTol',1e-4);
[t,xb]=ode45(Chua,[0 100],[-1.6,0,1.6],options);
plot3(xb(:,1),xb(:,2),xb(:,3))
title('Chua`s Double Scroll Attractor')

fsize=15;
xlabel('x(t)','Fontsize',fsize);
ylabel('y(t)','Fontsize',fsize);
zlabel('z(t)','FontSize',fsize);
```

Simulation Result :

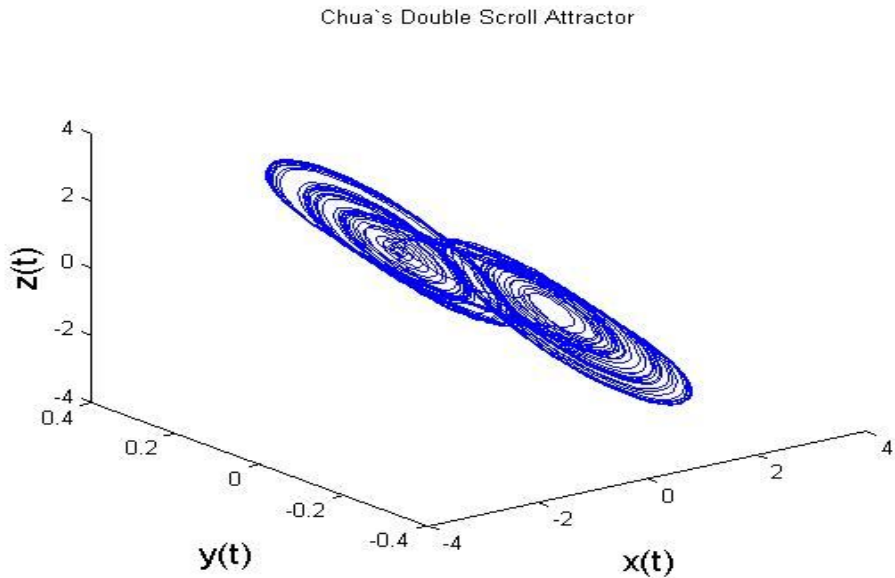

## (E) The Mandelbrot Set :

MATLAB Source code :

```
Nmax = 50;   scale = 0.005;
xmin = -2.4; xmax  = 1.2;
ymin = -1.5; ymax  = 1.5;

% Generate X and Y coordinates and Z complex values
[x,y]=meshgrid(xmin:scale:xmax, ymin:scale:ymax);
z = x+1i*y;

% Generate w accumulation matrix and k counting matrix
w = zeros(size(z));
k = zeros(size(z));

% Start off with the first step ...
N = 0;

% While N is less than Nmax and any k's are left as 0
while N<Nmax && ~all(k(:))
    % Square w, add z
    w = w.^2+z;
    % Increment iteration count
    N = N+1;
    % Any k locations for which abs(w)>4 at this iteration and no
    % previous iteration get assigned the value of N
    k(~k & abs(w)>4) = N;
```

```matlab
end

% If any k's are equal to 0 (i.e. the corresponding w's never blew up) set
% them to the final iteration number
k(k==0) = Nmax;

% Open a new figure
figure

% Display the matrix as a surface
s=pcolor(x,y,k);

% If you truly want the Mandelbrot curve in B&W, comment the above line and
% uncomment these two
% s = pcolor(x, y, mod(k, 2));
% colormap([0 0 0;1 1 1])

% Turn off the edges of the surface (because the cells are so small, the
% edges would drown out any useful information if we left them black)
set(s,'edgecolor','none')

% Adjust axis limits, ticks, and tick labels
axis([xmin xmax -ymax ymax])
fontsize=15;
set(gca,'XTick',xmin:0.4:xmax,'FontSize',fontsize)
set(gca,'YTick',-ymax:0.5:ymax,'FontSize',fontsize)
xlabel('Re z','FontSize',fontsize)
ylabel('Im z','FontSize',fontsize)

keyboard
figure
```

Simulation Result :

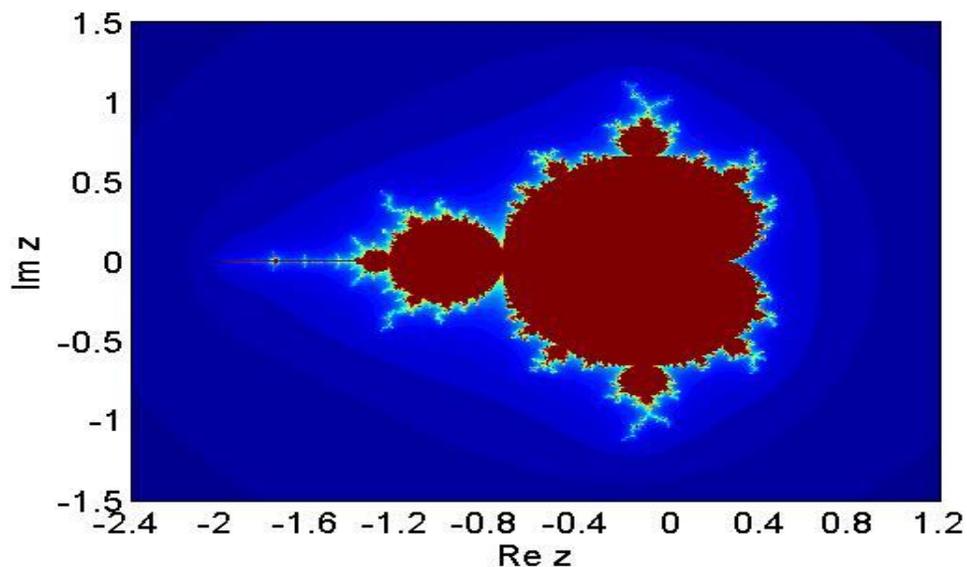

**6. Conclusion :** In this work, some basic applications of Dynamical Systems have been explored. Recently extensive research works in various other fields of Engineering have been going only to better understand these Dynamical Systems. Fractals have been successfully used in different applications like Image watermarking, De-noising, coding, indexing, and time series prediction. Fractal dimension idea can also be extended to distinguish between natural and artificial objects. Also fractals can be used in EEG signal analysis and fault classification. On the other side chaos can also be used in signal masking, and in random number generating. Bifurcation Analysis has also become an important analysis tool in stability studies of power system networks. Due to limited time, this study didn't go through those ongoing applications but definitely in upcoming days other useful Engineering applications of Dynamical Systems will emerge through research and studies like these.